\documentclass[a4paper]{mem}
\usepackage{natbib}
\usepackage{graphicx}
\usepackage[a4paper]{hyperref}
\idline{73}{23}
\begin{document}
\title{The Chemo-Dynamical Evolution of Elliptical Galaxies:
Pre-heating and AGN heating
}

\author{Daisuke Kawata, \and Brad K.\ Gibson\fnmsep
}

\offprints{D. Kawata}
\mail{Mail \#31,
PO Box 218, Hawthorn, VIC, 3122, Australia}

\institute{Centre for Astrophysics \& Supercomputing, 
  Swinburne University, Australia\\
 \email{dkawata,bgibson@astro.swin.edu.au}
  }

\abstract{
We study the chemodynamical evolution of elliptical galaxies and
their X-ray and optical properties using high-resolution cosmological
simulations. Our Tree N-body/SPH code includes a self-consistent treatment of
radiative cooling, star
formation, supernovae feedback, and chemical enrichment. We present a
series of ${\rm \Lambda}$CDM cosmological simulations which trace the
spatial and temporal evolution of abundances of heavy elements in both the
stellar and gas components of galaxies. 
A giant elliptical galaxy formed in one of the simulations
is quantitatively compared with the observational data 
in both the X-ray and optical regime.
We implement a treatment of both
pre-heating and AGN heating in this simulation, 
and examine the effect of these processes on elliptical galaxy formation.
We find that the adopted pre-heating ($T=10^7$ K at $z=4$)
is not strong enough to explain the observed X-ray or optical properties.
On the other hand, our AGN heating model in which the gas inflow induces
the AGN heating is consistent with both the X-ray and optical properties.
\keywords{galaxies: elliptical and lenticular, cD
---galaxies: formation---galaxies: evolution
---galaxies: stellar content
}
}
\authorrunning{D.\ Kawata and B.K.\ Gibson}
\titlerunning{The Chemo-Dynamical Evolution of Galaxies}
\maketitle
%

\section{Introduction}

 Developments in both ground- and space-based observational facilities
has made it possible for astronomers to understand
detailed self-consistent optical, X-ray, and theoretical
analyses of galaxy formation and evolution. For the case of
elliptical galaxies, optical observations provide constraints on 
the properties of the stellar component, while
X-ray observations constrain the physical conditions
of the hot interstellar medium (ISM). 

 \citet{kg03b} (KB03b) presented a first attempt
to explain both the X-ray and optical properties of observed elliptical
galaxies via the use of self-consistent cosmological simulations.
Using a standard ``recipe'' for galaxy formation,
they found that radiative cooling is important to interpret the
observed X-ray luminosity (${\rm L_X}$), temperature (${\rm T_X}$),
and metallicity (${\rm [Fe/H]_X}$) of the hot gas of elliptical galaxies.
However, they also found a serious problem in that the cooled gas also 
leads to excessive star formation at low redshift, and therefore 
results in underlying galactic
stellar populations which are too blue with respect to observations.
Their study demonstrates that the cross-check over both 
X-ray and optical properties 
provides stronger constraints for the theoretical models
(more so than anyone taken in isolation), and is essential
for any successful scenario of elliptical galaxy formation and
evolution. 

KG03b suggested that a heating mechanism which was not included
in their simulations would be required to
suppress this enhanced cooling and consequent
star formation.
In this paper, we examine two such potential mechanisms:
pre-heating and heating
by an Active Galactic Nucleus (AGN). Previous analytical
and numerical studies have shown that the pre-heating
of gas at high redshift may help to 
explain the X-ray luminosity and temperature of the hot gas
\citep[e.g.][]{bgw02}, 
although the source of the pre-heating remains unclear.
On the other hand, recent observations suggest that all 
elliptical galaxies have a central black hole \citep[e.g.][]{fm00}.
One might expect therefore that once the cold
gas falls into the central part of a galaxy,
the AGN may become ``active'' and heat up the
surrounding gas, and might be able to blow out the gas 
from the system via a radio jet.
Once the gas accretion is halted by such AGN heating, the AGN itself
loses its fuel ``source'' and goes back to a quiescent state, 
which allows the gas to cool again. This self-regulation
cycle may be capable of suppressing star formation
for a significant period of time, 
without changing the effect of radiative cooling,
which is in turn required to consider the 
indication for the observed X-ray properties.
Several earlier studies of the effect of AGN heating on
the hot gas of elliptical galaxies can be seen in \citet{bt95,bm02,bk02}.
Our current study examines the effect of pre-heating and AGN heating 
on the optical properties as well as the X-ray properties
in the context of a more self-consistent treatment of cosmological evolution.

\begin{figure*}
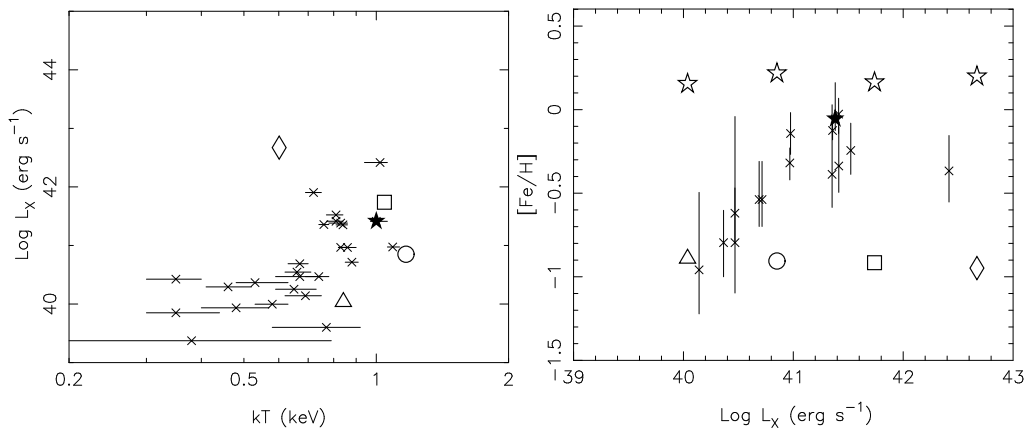

\centering
\resizebox{\hsize}{!}{\includegraphics[clip=true]{kgf1a.ps}
\includegraphics[clip=true]{kgf1b.ps}}
\caption{ Comparison of the simulated and observed
(crosses with error bars) ${\rm L_X-T_X}$ relations ({\it left panel})
and  ${\rm [Fe/H]_X-L_X}$ relations ({\it right panel}).
The circle/square/diamond/triangle indicates the predictions of 
Model 1/2/3/4. Open stars in the right panel
show the predicted mean metallicity of the stellar component
for the four models. The observational data are from \citet{mom00}.
The solid star presents the position of NGC 4472.
}
\label{lx-fig}
\end{figure*}

\section{Methods}

Our simulations were carried out using {\tt GCD+}, our original
galactic chemodynamical evolution code.
Details of the code are presented in \citet{dk99} and
\citet{kg03a,kg03b}.
In {\tt GCD+}, the dynamics of collisionless dark
matter and stars is calculated using a gravitational Tree N-body code,
and the gas component is modeled using Smoothed Particle Hydrodynamics
(SPH).  We calculate radiative cooling, star formation, chemical enrichment, 
and supernovae (SNe) feedback, self-consistently, and take into account both 
Type~Ia and Type~II SNe.  
We assume that SNe feedback is released as thermal energy.

 We have carried out a series of high-resolution
simulations within a standard $\Lambda$CDM cosmology
($\Omega_0$=0.3, $\Lambda_0$=0.7, 
$\Omega_{\rm b}$=0.019$h^{-2}$, $h$=0.7, and $\sigma_8$=0.9). 
Gas dynamics and star formation are included only within the relevant
high-resolution region
($\sim$12~Mpc at $z$=0); the surrounding low-resolution region ($\sim$43~Mpc)
contributes to the high-resolution region only through gravity.
The mass of individual gas particles in the high-resolution region 
was $5.9\times10^7$~${\rm M}_\odot$.  We identified an appropriate
elliptical galaxy analog in the high-resolution region, which acts as
the focus for this study.
The total virial mass of this target galaxy is
$2\times10^{13}$~${\rm M_\odot}$, similar in size to that of
NGC~4472, a bright
elliptical galaxy in the Virgo Cluster.  The target galaxy is relatively
isolated, with only a few low-mass satellites remaining at $z$=0.
This target galaxy is the same system discussed
in KG03b. Fig.\ 1 of KG03b shows the morphological evolution of 
dark matter in the simulation volume, and the evolution of the stellar
component of the target galaxy.
The galaxy forms through conventional hierarchical clustering
between redshifts $z$=3 and $z$=1; the morphology has not changed dramatically
since $z$=1. KG03b presented the results of three different radiative 
cooling and SNe feedback models. We use two of their models as reference
models for this study; Model 1 here corresponds to Model~B of KG03b, 
which includes cooling and weak feedback;
Model 2 here corresponds to Model C of KG03b 
which mimics Model 1, but incorporates
100 times greater thermal energy ($10^{52}$ erg) per supernova than
Model 1. Since in this study we are interested in the effect of the
pre-heating and the heating by AGN, we construct the additional
two models:

 Model 3: pre-heating model. The gas component in the 
entire simulation region is heated to a temperature of $T=10^7$ K
at redshift $z=4$. Weak thermal SNe feedback (as in Model 1)
is adopted.

 Model 4: AGN heating model. The most bounded 
star particle in the target galaxy at $z=1$ is assumed to be
the heating source, i.e.\ the AGN ``particle''. Since 
the AGN heating requires gas fueling, we assume that
the AGN heating is active only when the divergence of the velocity field
of the surrounding gas of the AGN particle is negative, 
$\langle {\bf \nabla} \cdot \mbox{\boldmath$v$}\rangle<0$.
The divergence of the velocity field
of the neighbour gas particles is calculated using the SPH scheme.
A constant thermal energy of $10^{44} {\rm erg\ s}^{-1}$ 
is deposited to the neighbour gas particles when the above condition
is satisfied. This energy is roughly consistent with observational
estimates \citep[e.g.][]{bsm03}.
Strong thermal SNe feedback (as in Model 2) is also adopted.

For all the models, we examine both the resulting X-ray {\it and}
optical properties, comparing them quantitatively with observation
(see KG03b for details).
The gas particles in our simulations carry with them knowledge of the 
density, temperature, and abundances of various heavy elements.
Using the XSPEC {\tt vmekal} plasma model, we derive the X-ray spectrum
for each gas particle, and synthesize them within the assumed aperture
(R$\sim$35~kpc). 
We next generate ``fake'' spectra with the response function of the XMM EPN
detector, assuming an exposure time (40~ks) and target galaxy distance
(17~Mpc).  Finally, our XPSEC fitting provides the X-ray weighted
temperatures and abundances of various elements.  Conversely, the
simulated star particles each carry their own age and metallicity ``tag'',
which enables us to
generate an optical-to-near infrared spectral energy distribution for
the target galaxy, when combined with our population synthesis code
adopting simple stellar populations of \citet{ka97}.

\begin{figure}
\centering
\includegraphics[width=\hsize]{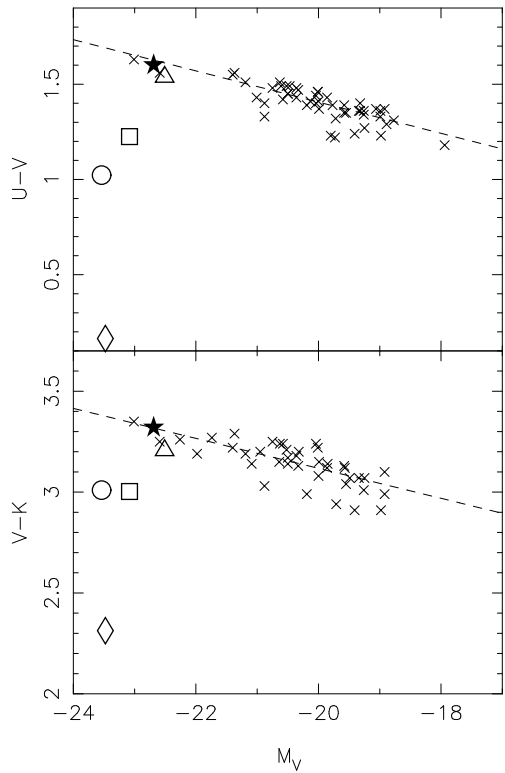}
\caption{
Comparison of the simulated $U-V$ and $V-K$ CMRs
(circle/square/diamond/triangle for Model 1/2/3/4) and
those of the Coma cluster ellipticals (crosses).
The observational data are from \citet{ble92}.
The dashed line shows the CMR fitted to the Coma Cluster galaxies.
The solid star shows the position of NGC 4472.
}
\label{cmr-fig}
\end{figure}


\begin{figure}
\centering
\includegraphics[width=\hsize]{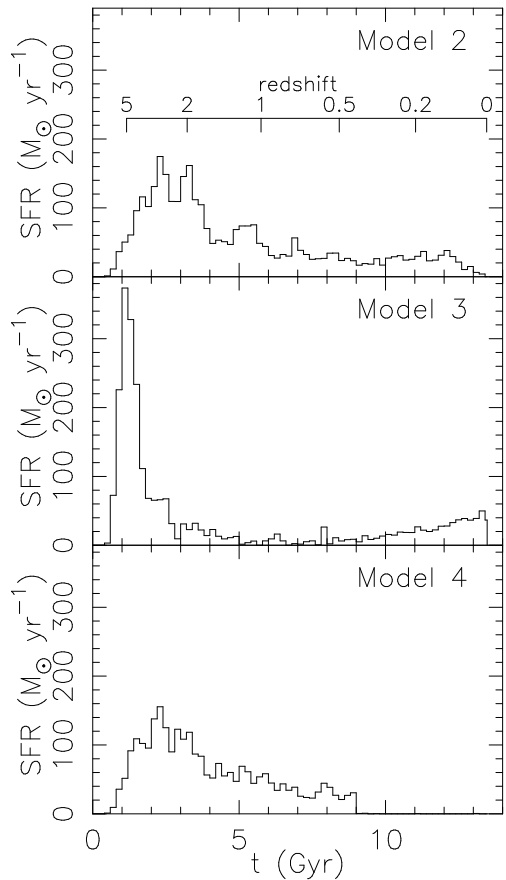}
\caption{
Time variation of the star formation rate
for Models 2 (upper), 3 (middle), and 4 (lower).
}
\label{sfr-fig}
\end{figure}

\begin{figure}
\centering
\includegraphics[width=\hsize]{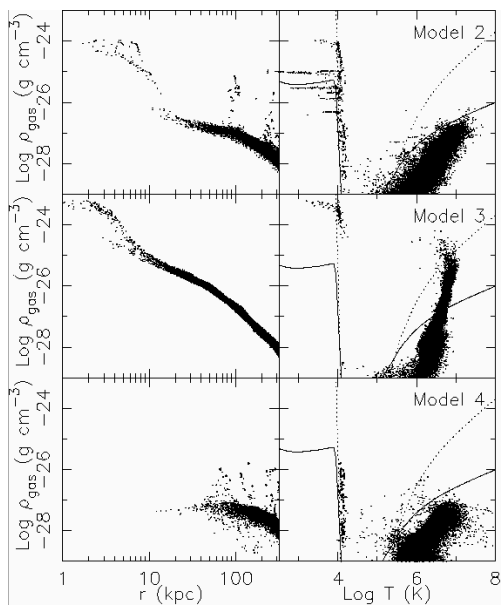}
\caption{Density vs radius (left) and density vs
temperature (right) distributions of gas particles 
for Models 1 (upper),  2 (middle), and 3 (lower). 
The solid (dotted) curves separate the
region where the cooling time is shorter (upper region) and longer
than the Hubble (dynamical) time.}
\label{tcdyn-fig}
\end{figure}

\section{Results and Conclusion} 

 To study both X-ray and optical properties, we examine
the ${\rm L_X-T_X}$ and ${\rm L_X-[Fe/H]_X}$ relations for the X-ray
properties, and the colour-magnitude relation (CMR) 
for the optical properties.
Figs.\ \ref{lx-fig} and \ref{cmr-fig} show 
the ${\rm L_X-T_X}$ and ${\rm L_X-[Fe/H]_X}$ relations 
and the CMRs for the four models, and compare the model results with
the observational data. Models 1 and 2 are re-plot from Models B and C
of KG03b. As shown in KG03b, current standard galaxy formation models,
such as Models 1 and 2, can roughly reproduce the ${\rm L_X-T_X}$ 
and ${\rm L_X-[Fe/H]_X}$ relations. 
However, the optical colours of the resulting stellar
components for these models
are inconsistent with the observational data (Fig.\ \ref{cmr-fig}), due to
excessive star formation at low redshift (Fig.\ \ref{sfr-fig}),
which is induced by radiative cooling.

 The pre-heating model, Model 3, leads to too high X-ray luminosity 
and too low temperature (Fig.\ \ref{lx-fig}).
Fig.\ \ref{tcdyn-fig} shows that in Model 2 the gas whose cooling
time, $t_{\rm cool}$, is shorter than the Hubble time, $t_0$,
is changed into the cold gas. However, in Model 3 the gas whose
$t_{\rm cool}$ is longer than the local
dynamical time, $t_{\rm dyn}=(3\pi/16G\rho)^{1/2}$,
can stay in the hot phase, which leads to the higher hot gas density,
and thus the higher ${\rm L_X}$. Also Fig.\ \ref{sfr-fig} shows that
the pre-heating cannot suppress star formation until $z=0$, although
star formation is suppressed until $z\sim0.5$ \citep[see also][]{tbs03}.
Consequently, a larger  
amount of remaining gas leads to a higher star formation rate at $z=0$,
which results in a bluer colour than Models 1 and 2.
This result demonstrates that the pre-heating we assumed here
is not strong enough for this elliptical galaxy.
A higher amount of the pre-heating will be examined in
a future work \citep[see also][]{bgw02}.

 The most successful model is the AGN heating model, Model 4. 
Although the X-ray luminosity is slightly lower 
compared with the observed value of NGC 4472 (Fig.\ \ref{lx-fig}),
Model 4 roughly reproduces both the X-ray and optical observational data.
Fig.\ 4 shows that our AGN heating model gets rid of the cold gas in
the central region, but does not change the hot gas profile
significantly. As a result, the hot gas properties for Model 4
are similar to Model 2.
This is due to the self-regulated AGN heating
which guarantees that the hot gas is not ``over-heated''.
We also found that, compared with Model 2, 
a larger fraction of iron ejected from stars 
is blown out from the system by the AGN heating, which results
in a lower metallicity in the hot gas (Fig.\ \ref{lx-fig}).
In addition, this AGN heating is powerful enough to stop the recent star 
formation, and leads to the observed red colours of elliptical
galaxies (Fig.\ \ref{cmr-fig}). 


Our cosmological chemodynamical code makes it possible to undertake
quantitative comparisons between numerical simulations
and observational data in both the X-ray and optical regime
with minimal assumptions.
We find that the AGN heating induced by cold gas inflow
can explain both the X-ray and optical observations of elliptical
galaxies. Although our AGN model is rather simple and
we focus on only one elliptical galaxy model, 
this result encourages more serious studies of
the effect of the AGN heating on elliptical galaxy formation.

\begin{acknowledgements}
We acknowledge the financial support of 
the Australian Research Council and the Japan Society for the Promotion 
of Science
through the Grants-in-Aid for Scientific Research (No.\ 14540221).
\end{acknowledgements}

\bibliographystyle{aa}

\begin{thebibliography}{}
\bibitem[Binney \& Tabor (1995)]{bt95}
 Binney J.J.\ \& Tabor G.\ 1995, MNRAS 276, 663
\bibitem[Blanton, Sarazin \& McNamara (2003)]{bsm03}
 Blanton E.L., Sarazin C.L., McNamara B.R.\ 2003, ApJ 585, 227
\bibitem[Bower, Lucey, \& Ellis (1992)]{ble92}
 Bower, R.G., Lucey, J..R., \& Ellis, R.S., 1992, MNRAS, 254, 589
\bibitem[Borgani et al.\ (2002)]{bgw02}
 Borgani S. et al.\ 2002, MNRAS 336, 409 
\bibitem[Brighenti \& Mathews (2002)]{bm02}
 Brighenti F.\ \& Mathews W.G.\ 2002 ApJL, 574, L11
\bibitem[Br\"uggen \& Kaiser (2002)]{bk02}
 Br\"uggen M., Kaiser C.R.\ 2002, Nature 418, 301
\bibitem[Ferrarese \& Merritt (2000)]{fm00}
 Ferrarese, L.\ \& Merritt, D., 2000, ApJL 539, 9
\bibitem[Kaiser (1991)]{nk91}
 Kaiser N.\ 1991, ApJ 383, 104
\bibitem[Kawata (1999)]{dk99}
 Kawata D.\ 1999, PASJ 51, 931
\bibitem[Kawata \& Gibson (2003a)]{kg03a}
 Kawata D.\ \& Gibson B.K.\ 2003a, MNRAS 340, 908
\bibitem[Kawata \& Gibson (2003b)]{kg03b}
 Kawata D.\ \& Gibson B.K.\ 2003b, MNRAS 346, 135 (KG03b)
\bibitem[Kodama \& Arimoto (1997)]{ka97}
 Kodama T.\ \& Arimoto N.\ 1997 A\&A, 320, 41 
\bibitem[Matsushita, Ohashi, \& Makishima (2000)]{mom00}
 Matsushita, K., Ohashi, T., \& Makishima, K., 2000, PASJ, 52, 685
\bibitem[Tornatore et al.\ (2003)]{tbs03}
 Tornatore L.\ et al.\ 2003, MNRAS 342, 1025
\end{thebibliography}

\end{document}